**Morphology Characterization of Argon-Mediated Epitaxial Graphene on C-face SiC**

J.L. Tedesco[1], G.G. Jernigan[2], J.C. Culbertson[2], J.K. Hite[1], Y. Yang[1], K.M. Daniels[1], R.L. Myers-Ward[1], C.R. Eddy, Jr.[1], J.A. Robinson[3], K.A. Trumbull[3], M.T. Wetherington[3], P.M. Campbell[2], and D.K. Gaskill[1]

[1] Advanced Silicon Carbide Epitaxial Research Laboratory, U.S. Naval Research Laboratory, 4555 Overlook Avenue SW, Washington, D.C. 20375
[2] U.S. Naval Research Laboratory, 4555 Overlook Avenue SW, Washington, D.C. 20375
[3] Electro-Optics Center, The Pennsylvania State University, 559A Freeport Road, Freeport, PA 16229

**Abstract**

Epitaxial graphene layers were grown on the C-face of 4H- and 6H-SiC using an argon-mediated growth process. Variations in growth temperature and pressure were found to dramatically affect the morphological properties of the layers. The presence of argon during growth slowed the rate of graphene formation on the C-face and led to the observation of islanding. The similarity in the morphology of the islands and continuous films indicated that island nucleation and coalescence is the growth mechanism for C-face graphene.

Epitaxial graphene grown on SiC has been shown to exhibit a wide range in carrier mobility, with reports as high as 20,000 cm$^2$V$^{-1}$s$^{-1}$ [1-6]. It is clear that the method of graphene production can have a dramatic affect on the measured mobility [5]. Recent reports have indicated that an argon (Ar) overpressure used during graphene growth can result in improved morphology and mobility [7-10]. This work provides a more complete study of graphene growth on C-face substrates of 4H- and 6H-SiC polytypes as a function of Ar pressure, from 10$^{-5}$ mbar (*in vacuo*) to 200 mbar, and temperature, from 1,200°C to 1,700°C, in the same induction-heated chemical vapor deposition (CVD) reactor to develop an understanding of graphene formation in the presence of Ar. For graphene growth on C-face substrates, it was found that the presence of Ar slowed the growth rate of graphene, as shown by an intermediate island formation stage and by higher growth temperatures required for complete film coverage. As the pressure of Ar increased, the graphene morphology became more uniform, most likely resulting from fewer island nucleation sites, and with more uniform graphene came an improvement in mobility [8-9].

The substrates used in this study were semi-insulating, on-axis (0° ± 0.5°), C-face 4H- (Cree) and 6H-SiC (II-VI, Inc.) wafers. The standard sample size was 16 × 16 mm$^2$ ("centimeter-scale") that had been sawn from the 76.2 mm diameter wafers. Full 50.8 mm diameter, semi-insulating, on-axis (0° ± 0.5°), Si-face 6H-SiC (II-VI, Inc.) wafers were also used as substrates for growth ("wafer-scale" samples). All samples were chemically cleaned *ex situ* using a standard SiC cleaning procedure developed in this laboratory [11] prior to loading into a commercial, inductively heated, hot wall CVD reactor (Aixtron/Epigress VP508). Following a 1,600°C hydrogen etch to remove polishing damage [3,12-13], epitaxial graphene was grown at temperatures from 1,200°C to 1,700°C for times ranging from 60 to 120 minutes and using Ar pressures of 50, 100, and 200 mbar. *In vacuo* growths (10$^{-5}$ mbar) utilized the standard high vacuum silicon sublimation method described elsewhere [3,6] while argon-mediated growths were performed using a procedure described previously [9]. Following growth, the samples were removed from the reactor, and graphene formation was confirmed using Raman spectroscopy, for the case of island formation [9], and additionally the finite resistance method [6], for the case of continuous film growth. Morphologies were evaluated using a commercial Nomarski interference microscope (Olympus BX60M) and a commercial tapping mode atomic force microscope (AFM: Digital Instruments Dimension 3100). The thickness uniformity was measured using 488 nm Raman topography mapping using a method reported previously [14-16].

Figure 1 shows a tapping mode atomic force microscopy (AFM) image and 3-dimensional representation of an isolated island formed on the C-face after a 60 minute long growth at 1,550°C in 100 mbar of Ar. Raman spectroscopy was used to confirm that the island was indeed graphene. The image illustrates a number of important features relevant to graphene growth on the C-face of SiC. "Ridges" (also referred to as "puckers" [17] or "giraffe stripes" [13]) were observed around the perimeter of the island and across the island. The ridges on the island appear the same as the ridges in continuous films [3,13,18]. It has been proposed that these ridges form as a strain relaxation mechanism during cool-down after growth, resulting from different thermal expansion coefficients between the graphene and the SiC [19]. This is quite possible, but the observation of ridges at the edge of the island, where the graphene is not constrained, indicates that there are additional phenomena occurring. It is likely that the ridges result from the graphene growing vertically [20] or the edge of the graphene curling like a nanotube. If this is the case, the center ridge may be the result of the impingement of two adjacent graphene platelets. Additionally, the graphene island is growing inside a depression of the SiC surface. In essence, carbon atoms become available to form graphene as Si is depleted nearby. Because approximately three bilayers of SiC are needed to decompose in order to provide enough carbon atoms to form one layer of graphene for this basal plane orientation [17], the depression in the SiC is deeper than the graphene island is tall (with the exception of the ridges). The SiC step edges also conform around the region of the growing graphene island. Hydrogen etching is performed prior to graphene growth and results in relatively uniform step-edges on the surface [13]. Hence, a bulk defect may have existed at this island location, which may have initiated the graphene growth, and would be responsible for the non-uniform SiC step morphology.

An example of the later stages of graphene island growth is shown in Fig. 2, where graphene was formed at 1,575 °C for 60 minutes in 100 mbar of Ar. The AFM image shows two large regions of graphene (as distinguished by the presence of ridges) and a region of bare SiC between them. The 3-dimensional representation shows how much rougher the graphene morphology has become compared to the smaller island shown in Fig. 1 and the surrounding SiC surface. The graphene region also has a larger depression relative to the SiC surface when compared to the smaller island. The similarity of the ridges between stages of island growth and final morphology implies that the formation of ridges is not related to the growth rate, but is intrinsic to C-face graphene.

Raman attenuation of the underlying SiC signal can be used to determine the thickness of graphene [21] growth on C-face SiC. Figure 3 shows an example of a Raman measurement on an island formed at 1600°C for 10 minutes at 100 mbar of Ar, along with a Nomarski image of the island. The thickness of the graphene within the island varies ~7.5 nm from the center to the edge. This is typical for graphene islands observed on C-face SiC during growths under Ar. The graphene thickness is greater in the center of the islands than at the edges and optically observable. From this, it appears that graphene grows in a manner similar to a Volmer-Weber mechanism in deposited films [22]. It can be hypothesized that, as the growth proceeds, the C-face growth mechanism allows the films to become thicker as the islands deplete nearby SiC steps and that the mosaic structure of the ridges results from the junction of graphene islands as they coalesce. Given this growth mechanism, future studies would be warranted to determine whether increasingly longer growths would result in increasingly thick graphene films, or if eventually the C-face graphene reaches steady-state saturation, similar to what has been observed in graphene growth on the Si-face [5,23].

The Ar growth pressure and growth temperature range over which graphene islands can be observed on the C-face are shown in Fig. 4. As expected from the slower growth rate in Ar, higher growth temperatures are needed to achieve complete film growth as the Ar pressure is increased. Correspondingly, the temperature necessary to initiate growth also increases as the Ar pressure increases. The window over which islands were observed is ~100°C for each Ar pressure investigated. There is a temperature regime where *in vacuo* islands can be observed, therefore, islanding is not due solely to the fact that Ar suppresses graphene growth on the C-face. Raman spectra indicate that as the growth goes from islanding to complete films the ratio of the intensity of the D peak (~1,350 cm$^{-1}$) to the G peak (~1,570 cm$^{-1}$) decreases. It has been shown that this ratio is inversely proportional to crystallite size [24-25]. Averaging the Raman spectra taken over a large area (40 μm × 40 μm with a 1 μm step size), the SiC background was subtracted and the average D/G ratio was calculated. Using this method, the initial islands were found to have a range of crystallite dimensions between 100 nm and 200 nm, and as growth progresses, the final films develop crystallites with dimensions ranging from 1.5 to 2.0 μm. This data suggests that improvements in morphology in graphene growth performed under an Ar ambient arise from a limiting of the number of nucleation sites and subsequent island coalescence to produce larger domains. A significant increase in graphene film thickness once a complete coverage had developed. This is most likely the result of Si desorption from the surface becoming hindered by the graphene film.

Epitaxial graphene films were grown on the C-face of SiC in a CVD reactor using *in vacuo* and argon-mediated growth processes. It was observed that argon-mediated growth on the C-face could produce graphene islands, whose appearance was a function of temperature and Ar pressure. Conditions were found where islands grew during *in vacuo* growth implying that Ar does not change the growth mechanism. The fact that islands formed first for *in vacuo* as well as for argon-mediated graphene implied that island nucleation and coalescence is the mechanism for growth of C-face graphene consistent with a Volmer-Weber mechanism. Furthermore, ridges were identified on both islands and continuous films, implying that they are intrinsic to C-face epitaxial graphene, and that they are more complicated than being simply a strain relaxation mechanism following film growth.

**Acknowledgements**

The authors acknowledge the Office of Naval Research and Defense Advanced Research Projects Agency Carbon Electronics for RF Applications program for funding. J.L.T. and J.K.H. acknowledge the American Society for Engineering Education for support through Naval Research Laboratory Postdoctoral Fellowships. Y.Y. and K.M.D. acknowledge support from the Naval Research Enterprise Intern Program.

**Figures**

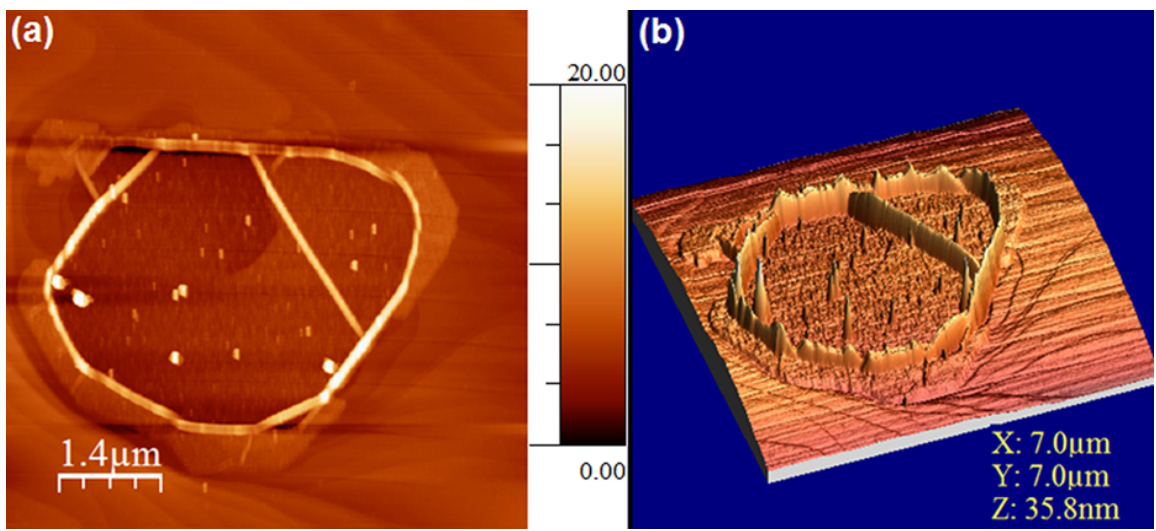

Fig. 1 (color online). (a) Atomic force microscopy image of a small graphene island and (b) three-dimensional representation of the island shown in (a) formed at 1,550°C in 100 mbar Ar for 60 minutes. The height scale for (a) is in units of nanometers.

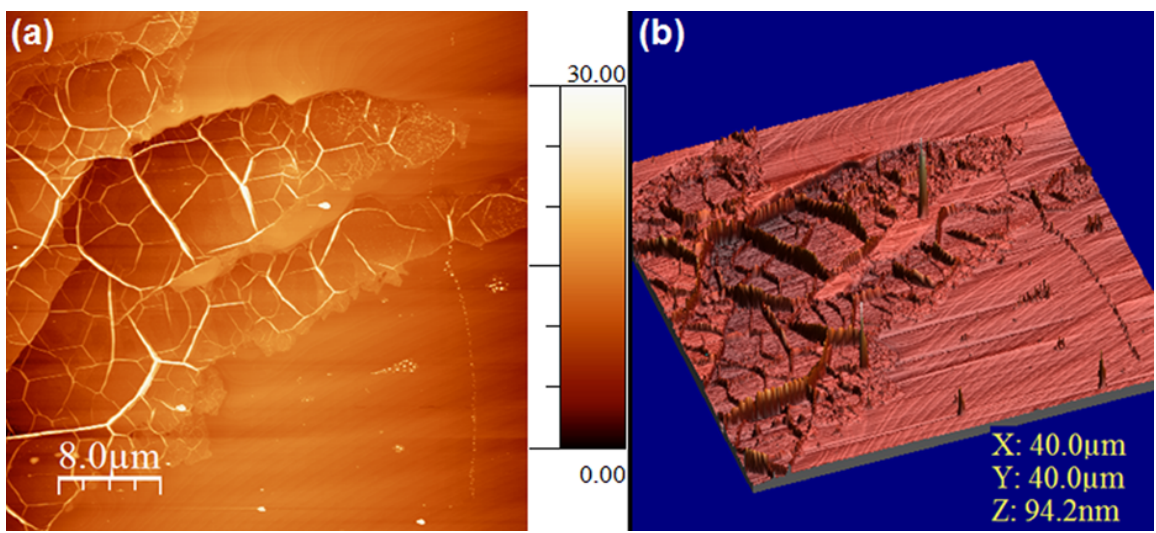

Fig. 2 (color online). (a) Atomic force microscopy image of a larger graphene island and (b) three-dimensional representation of the island shown in (a) formed at 1,575°C in 100 mbar Ar for 60 minutes. The height scale for (a) is in units of nanometers.

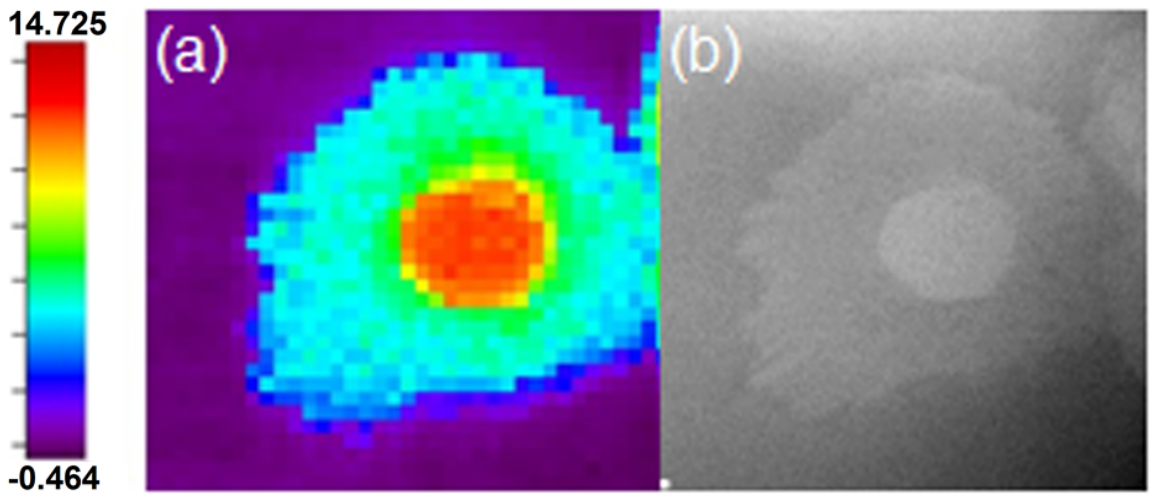

Fig. 3 (color online). (a) Raman thickness mapping and (b) Nomarski contrast image of a graphene island formed at 1600 °C in 100 mbar Ar for 10 minutes. The height scale for (a) is in units of nanometers.

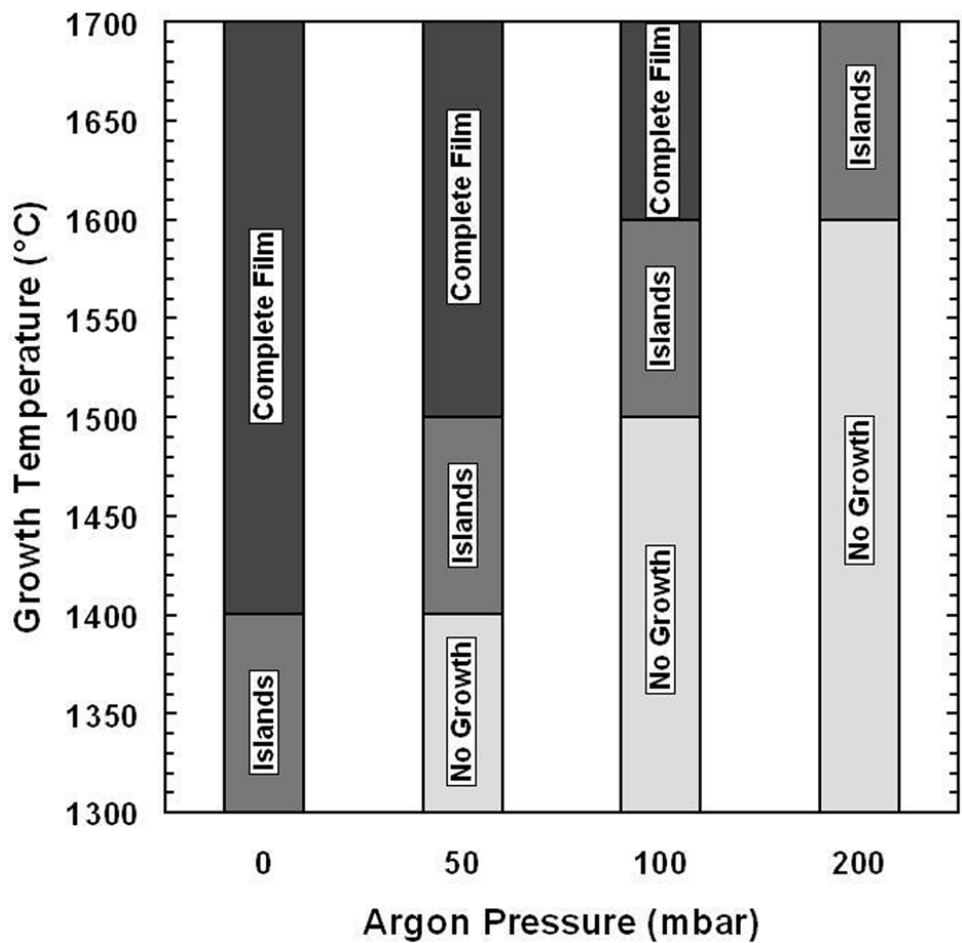

Fig. 4. General descriptions of the morphology of the epitaxial graphene as a function of growth temperature and Ar growth pressure. Zero Ar pressure corresponds to an *in vacuo* growth at $10^{-5}$ mbar.